\documentclass[useAMS,usenatbib]{mn2e}
\usepackage{epsfig}
\usepackage{color}
\usepackage{subfigure}
\usepackage{natbib}
\usepackage{aas_macros} 

\title[Cosmology from strong lenses]{Cosmological parameters from lenses distance ratio}

\author[V.F. Cardone et al.]{Vincenzo F. Cardone$^{1}$\thanks{Corresponding author\,: {\tt winnyenodrac@gmail.com}}, Ester Piedipalumbo$^{2,3}$, Paolo Scudellaro$^{2,3}$ \\
$^1$ INAF -- Osservatorio Astronomico di Roma, via Frascati 33, 00040 - Monte Porzio Catone (Roma), Italy \\
$^2$ Dipartimento di Scienze Fisiche, Universit\`{a} degli Studi di Napoli "Federico II", Compl. Univ. Monte S. Angelo, \\
\ \ Ed. 6, via Cinthia, 80126 Naples, Italy \\
$^3$ I.N.F.N., Sez. di Napoli, Compl. Univ. Monte S. Angelo, Ed. 6, via Cinthia, 80126 - Napoli, Italy}

\date{Accepted xxx, Received yyy, in original form zzz}

\begin{document}

\maketitle

\begin{abstract}

Strong lensing provides popular techniques to investigate the mass distribution of intermediate redshift galaxies, testing galaxy evolution and formation scenarios. It especially probes the background cosmic expansion, hence constraining cosmological parameters. The measurement of Einstein radii and central velocity dispersions indeed allows to trace the ratio $D_{s}/D_{ls}$ between the distance $D_{s}$ from the observer to the source and the distance $D_{ls}$ from the lens to the source. We present an improved method to explicitly include the two-component structure in the galaxy lens modeling, in order to analyze the role played by the redshift and the model dependence on a nuisance parameter,$ \ {\cal{F}}_E $, which is usually marginalized in the cosmological applications. We show how to deal with these problems and carry on a Fisher matrix analysis to infer the accuracy on cosmological parameters achieved by this method.

\end{abstract}


\section{Introduction}

Dozen years after its unexpected and somewhat serendipitous discoverty, the accelerated expansion of the universe is taken for granted due to the flood of data from different astrophysical probes confirming it (see, e.g., \cite{W12} for a not updated yet exhaustive review). Although the spatially flat concordance $\Lambda$CDM model, made out of a cosmological constant accounting for $\sim 70\%$ of the energy budget and responsible of the cosmic speed up, is in full agreement with observations (as summarized, e.g., in \cite{PlanckXIII}, where the latest datasets are considered), it is far from free of any conceptual and theoretical problems \citep{Carroll,MarekZC15}. This motivated the search for alternative models generally referred to as dark energy (DE) ones and characterized by the presence of a scalar field with a negative equation of state (EoS) denoted as $w = p/\rho$, with $(p, \rho)$ the pressure and energy density of this leading component. While a huge number of papers has addressed the problem of what DE is with proposal running from self interacting fields to modified gravity theories (see, e.g., \cite{AT15} for a recent textbook), observations are mainly aimed at constraining the DE EoS. Under the simple yet efficient CPL \citep{CP01,L03} parameterization, it is considered $w = w_0 + w_a (1 - a)$, with $a = 1/(1+z)$ the scale factor and $z$ the redshift, so that the Holy Grail of observational cosmology has nowadays become to narrow down as much as possible the range for the $(w_0, w_a)$ parameters. As a consequence, the efficacy of an observational probe is presently quantified in terms of the Figure of Merit (FoM) defined as the inverse of the area delimiting the $95\%$ confidence range in the $(w_0, w_a)$ space \citep{DETF}.

Gravitational lensing has soon been pointed at as one of the most promising tools to investigate the nature of the mysterious DE. In its journy from the source to the observer, the light is deflected by both a dominant mass concentration (a galaxy or a cluster acting as lens) along the line of sight and the cumulative effect of the large scale structure. As a consequence, gravitational lensing probe both the background expansion and the growth of structures. Cosmic shear tomography, that is to say the power spectrum of shear as measured from galaxies separated in different redshift bins \citep{BS01,HJ08}, has emerged as one of the most promising tools to constrain the DE EoS, so motivating the interest in ongoing (DES, KIDS) and future (Euclid, LSST) dedicated surveys. While shear tomography is based on the weak regime of gravitational lensing (where lensed galaxies are subject to tiny modifications detectable only on statistical grounds), also statistics based on strong lensing features is not less important. In this regime, the light emitted by the source passes close to a massive lens causing the formation of multiple images or spectacular Einstein rings or radial and/or tangential arcs. The distribution of angular separation between double images \citep{LO02,LM04,M05,Z09}, arc statistics \citep{Bart03,DHH04,Men13,B15} and Einstein rings properties \citep{BB08,Zit12,W14} have been used as tools to constrain cosmological parameters and the astrophysical properties of the lens population.

Galaxy scale strong lenses have been also proposed as a tool to probe the background expansion. The basic idea is  usually to infer the distance ratio $D_{ls}/D_s$ between the angular diameter distance from the lens to the source and from the observer to the source, and then use a sufficiently large sample to trace how this quantity evolves with $z$. To this end, one typically relies on  measurements of both the Einstein radius $R_E$ and the lens velocity dispersion $\sigma_0$. Under the assumption of isothermal sphere, one gets $R_E = 4 \pi (D_{ls}/D_s) (\sigma_{SIS}/c)^2$, so that a measurement of $(R_E, \sigma_{0})$ immediately gives the distance ratio $D_{ls}/D_s$ provided one sets $\sigma_{SIS} = f_E \sigma_0$, being $f_E$ a nuisance parameter to marginalize over \citep{Bie10,Cao12,Pior13}. Although quite simple, such a method however rests on some assumptions which,  even if reasonably motivated, should nonetheless be verified. First, it is assumed that all the lenses can be modelled as singular isothermal sphere. Second, the nuisance parameter $f_E$ is taken as a sort of universal constant, while its actual value can change from one system to another depending on both the lens properties (redshift, mass, size) and how $\sigma_0$ is measured (weighted within a circular aperture of fixed physical radius or through longslit spectroscopy). Here we follow a similar approach, but weakening the strong assumptions of the standard approach in order to avoid biasing cosmological parameters because of incorrect lens modelling. 

The plan of the paper is as follows. In Sect.\,2, a step\,-\,by\,-\,step derivation of the method is presented underlying which assumptions are made and how we deal with them. Sect.\,3 is devoted to validate the proposed technique showing that its basic assumptions are well motivated. We then present, in Sect.\,4,  a Fisher matrix analysis to forecast the errors on cosmological parameters for different configurations of future datasets. A summary of the results and a discussion of future perspectives are given in the concluding Sect.\,5. 

\section{Distance ratio}

Galaxy scale strong lenses are typically used to constrain the dark matter halo and its interplay with the properties of the stellar component. To this end, two observational probes are used, namely the projected mass $M_{proj}$ within the Einstein radius $R_E$ and the central velocity dispersion $\sigma_0$. Assuming the galaxy may be modelled as the sum of a stellar part modelled with a Sersic (1968) profile and a dark matter halo, we get\footnote{Unless otherwise stated, length quantities such as the Einstein radius are meant to be in linear units. When this is not the case, we will explicitly give the angular units adopted.}

\begin{equation}
M_{proj}^{th}(R_E) = M_{\star} \mu_{\star}(R_E) \left [ 1 + \frac{M_{vir}}{M_{\star}} {\mu_{DM}(R_E)}{\mu_{\star}(R_E)} \right ]\,, 
\label{eq: mprojth}
\end{equation}
with $(M_{\star}, M_{vir})$ the total stellar and virial masses, respectively, and $\mu(R)$ is a dimensionless function depending on the density profile. For the stellar component this reads

\begin{equation}
\mu_{\star}(R) = \frac{\Gamma(2n) - \Gamma[2n, b_n (R/R_{eff})^{1/n}]}{\Gamma(2n)}\,,
\label{eq: defmustar}
\end{equation}
with $n$ the slope of the Sersic model, $b_n$ a constant set so that half of the total luminosity is contained within the effective radius $R_{eff}$ \citep{Ciotti91}. We leave for the moment the $\mu_{DM}(R)$ function undetermined in order to be as general as possible.

Provided that the lens and source redshifts $(z_l, z_s)$ are known, one can get an estimate of the projected mass by simply measuring the Einstein radius and then using

\begin{equation}
M_{proj}^{obs}(R_E) = \pi \Sigma_{crit}(z_l, z_s) R_E^2 = \frac{c^2 R_E^2(arc)}{4G (206265)^2} \frac{D_{l} D_{s}}{D_{ls}}\,,
\label{eq: mprojobs}
\end{equation}
where we have used the definition of critical surface density $\Sigma_{crit}(z_l, z_s)$:

\begin{equation}
\Sigma_{crit} = \frac{c^2}{4 \pi G} \frac{D_{s}}{D_{l} D_{ls}}\,,
\label{eq: defsigmacrit}
\end{equation}
being $c$ the speed of light and $(D_{l}, D_{s}, D_{ls})$ the angular diameter distances to the lens, the source and between lens and source. Note that, in Eq.(\ref{eq: mprojobs}), $R_E(arc) = R_E (206265/D_{l})$ is the Einstein radius in arcsec since this is the quantity directly measured from the data in a model independent way.

The second quantity of interest is the aperture velocity dispersion, i.e. the velocity dispersion luminosity weighted within a circular aperture of radius $R_{ap}$. Approximating the deprojected Sersic profile with the \cite{PS97} model, it turns out that

\begin{eqnarray}
\sigma_{ap}^2 & = & \frac{G M_{\star}}{R_{eff}} \frac{b_n^{n(3  - p_n)} \Gamma(2n)}{n {\rm e}^{b_n} \Gamma[n(3  - p_n)]} \\
 & \times & \frac{{\cal{I}}_{\star}(R_{ap}; n) + (M_{vir}/M_{\star}) {\cal{I}}_{DM}(R_{ap}; n, R_{eff}; {\bf p}_{DM})}
{\Gamma(2n) - \Gamma[2n, b_n (R_{ap}/R_{eff})^{1/n}]} \nonumber \ ,
\label{eq: sigmaapth}
\end{eqnarray}
where $p_n$ is given in \cite{Metal01}, ${\bf p}_{DM}$ collectively denotes the halo model parameters, and we refer to \cite{C09} and \cite{CT10} for the details of the dimensionless ${\cal{I}}(x, {\bf p})$ functions. For dimensional reasons it is convenient to define an aperture mass as 

\begin{eqnarray}
M_{ap}^{th} & = & \frac{R_{eff} \sigma_{ap}^2}{G} =
 M_{\star} \frac{b_n^{n(3  - p_n)} \Gamma(2n)}{n {\rm e}^{b_n} \Gamma[n(3  - p_n)]} \\
 & \times & \frac{{\cal{I}}_{\star}(R_{ap}; n) + (M_{vir}/M_{\star}) {\cal{I}}_{DM}(R_{ap}; n, R_{eff}; {\bf p}_{DM})}
{\Gamma(2n) - \Gamma[2n, b_n (R_{ap}/R_{eff})^{1/n}]} \nonumber \ ,
\label{eq: mapth}
\end{eqnarray} 
which can be straightforwardly estimated from measurable quantities as

\begin{equation}
M_{ap}^{obs} = \frac{R_{eff}(arc)}{206265} \frac{D_l \sigma_{ap}^2}{G}\,,
\label{eq: mapobs}
\end{equation}
with $R_{eff}(arc)$ the effective radius in arcsec obtained from the fit to the surface brightness profile.

It is now only a matter of algebra to first solve $M_{ap}^{obs} = M_{ap}^{th}$ with respect to $M_{\star}$ and then insert the result into $M_{proj}^{obs}(R_E) = M_{proj}^{th}(R_E)$ to finally get

\begin{equation}
{\cal{D}}(z_l, z_s) = {\cal{D}}_{obs} \ \times \ {\cal{F}}_E \ ,
\label{eq: testeq}
\end{equation}
which represents the basic relation of our method relating the theoretical distance ratio 
\begin{equation}
{\cal{D}}(z_l, z_s) = D_{s}/D_{ls} 
\label{eq: defdzlzs}
\end{equation}
on the left hand side with the $({\cal{D}}_{obs}, {\cal{F}}_E)$ quantities on the right hand side defined as follows
 
\begin{eqnarray}
{\cal{D}}_{obs} & = & \frac{4 \times 206265}{R_E(arc)} \frac{R_{eff}(arc)}{R_E(arc)} \left ( \frac{\sigma_{ap}}{c} \right )^2 
\frac{n {\rm e}^{b_n} \Gamma[n(3 - p_n)]}{b_n^{n(3 - p_n)}} \nonumber \\
 & \times & \left \{ 1 - \frac{\Gamma[2n, b_n (R_{ap}/R_{eff})^{1/n}]}{\Gamma(2n)} \right \} \nonumber \\
 & \times & \left \{1 - \frac{\Gamma[n(3 - p_n), b_n (R_{ap}/R_{eff})^{1/n}]}{\Gamma[n(3 - p_n)]} \right \} \ ,
\label{eq: defdobs}
\end{eqnarray}

\begin{equation}
{\cal{F}}_E = \frac{1 + (M_{vir}/M_{\star}) [\mu_{DM}(R_E; {\bf p}_{DM})/\mu_{\star}(R_E; n)]}
{{\cal{I}}_{\star}(R_{ap}; n) + (M_{vir}/M_{\star}) {\cal{I}}_{DM}(R_{ap}; n, R_{eff}; {\bf p}_{DM})} \ .
\label{eq: defcalfe}
\end{equation}
Eq.(\ref{eq: testeq}) is formally similar to the one used in previous similar methods expressing the distance ratio in terms of a quantity depending only on measurable quantities (given by ${\cal{D}}_{obs}$) and an unknown nuisance parameter (referred to as ${\cal{F}}_E$ here). However, we improve on the classical approach in two ways. First, we have not assumed the galaxy to be a one component system, but rather modelled it as it actually is, i.e. the sum of a stellar part\footnote{By adopting a single Sersic profile, we are implicitly assuming that the lens is an early\,-\,type galaxy, otherwise one should split the stellar term in the sum of a Sersic bulge and an exponential disk. This is not a serious limitation since most of the lenses are indeed intermediate ellipticals.} and a dark matter halo. Second, we have now an analytical expression for the nuisance parameter, so that we can investigate whether it is the same for all lenses or rather it is a function of both the lens properties and the halo model.

\section{The nuisance parameter}

To infer the distance ratio from strong lensing data, one generally assumes that the $f_E$ parameter is a constant independent on the lens and marginalizes over it. This is actually a zero order approximation related to the properties of the assumed singular isothermal sphere model used to describe the total mass density. On the contrary,  with our ${\cal{F}}_E$ parameter in Eq.(\ref{eq: defcalfe}) we account for  the two components nature of the lens and explicitily specify the halo model. 

A careful analysis of the different terms entering its definition shows that ${\cal{F}}_E$ depends on  (i) the mass ratio $M_{vir}/M_{\star}$, (ii) the length ratio $R_{-2}/R_{eff}$ (with $R_{-2}$ the radius where the DM logarithmic slope equals -2), and (iii) the halo concentration\footnote{No matter which halo model is adopted, it is always possible to define the concentration $c_{vir}$ provided that the logarithmic slope is not constantly equal to -2, which is the case for the singular isothermal sphere only.} $c_{vir} = R_{vir}/R_{-2}$. Actually, the halo model can be more complex, so that $(M_{vir}, c_{vir})$ are not enough to fully assign it, and more parameters are needed, which makes ${\cal{F}}_E$ depend on them too. Moreover, since  the mass and length ratios may both change with the redshift as it is suggested by, e.g., the redshift dependence of the $c_{vir}$\,-\,$M_{vir}$ relation as well as of the $R_{eff}$\,-\,$M_{\star}$ scaling law, it turns out that ${\cal{F}}_E$ is redshift dependent.

These considerations strongly point against the standard assumption to consider the nuisance parameter as a constant, which implies that one should fit for ${\cal{F}}_E$ separately for each lens, hence making the method unusable. However, a possible way out could be found if ${\cal{F}}_E$ was constant within a certain scatter once the lenses were binned according to some suitably chosen properties. To investigate whether this is possible or not, we proceed as schematically pointed out below.

\begin{enumerate}

\item{We set the lens redshift $z_l$ sampling from a uniform distribution over the range $(0.1, 2.1)$ and then set $z_s = \zeta z_l$ with $\zeta$ randomly extracted from the range $(1.1, 3.1)$.} \\

\item{We use the redshift dependent galactic stellar mass function used in \cite{Fontana} to set the lens stellar mass $M_{\star}$ and sample the effective radius $R_{eff}$ from a Gaussian distribution centred on $R_{ref}(M_{\star}, z = 0) (1 + z)^{\alpha}$ and a $10\%$ scatter.  The estimation of  $R_{ref}(M_{\star}, z = 0)$ is obtained following \cite{Shen}, while the scaling with redshift has been chosen in accordance with \cite{Trujillo} considering the value of $\alpha$ for massive galaxies ($M_{\star} > 3 \times 10^{10} \ {\rm M}_{\odot}$).} \\

\item{The halo virial mass is set solving the $M_{\star}/M_{vir}$ vs $M_{vir}$ relation of \cite{Moster10} with respect to $M_{vir}$, so that we can finally fix the concentration sampling from the $c_{vir}$\,-\,$M_{vir}$ relation of \cite{Duffy08}, taking also into account its  scatter.} \\

\item{Adopting an NFW \citep{NFW96,NFW97} model for the DM halo, we can now compute both the Einstein radius $R_E$ and the central velocity dispersion $\sigma_0 = \sigma_{ap}(R_{ap} = R_{eff}/8)$. We then retain the simulated lens only if its input and output quantities are reasonable. To check whether this is the case, we compare them to the similar quantities for the SLACS \citep{Slacs} sample of 85 intermediate redshift lenses with well measured values of $(z_l, z_s, M_{\star}, R_E, \sigma_0)$.}
\end{enumerate}

Using the above recipe, we can generate a sample of $\approx 10000 $ lenses, with a distribution in $(z_l, z_s, M_{\star}, R_E, \sigma_0)$ similar to those of the real SLACS, and confidently use it to investigate how the nuisance parameter, ${\cal{F}}_E$, scales with the lens properties. Moreover, since we are interested more in the distance ratio estimate than in ${\cal{F}}_E$,  we wonder what  could be the bias on  ${\cal{D}}(z_l, z_s)$ caused by using  the mean value ${\cal{F}}_E = \langle {\cal{F}}_E \rangle$, averaged over a subsample of 1000 lenses selected according to a given criterion, instead of ${\cal{F}}_E$. To this end, we define the distance ratio estimator

\begin{figure*}
\resizebox{\hsize}{!}
{\includegraphics[width=3.5cm]{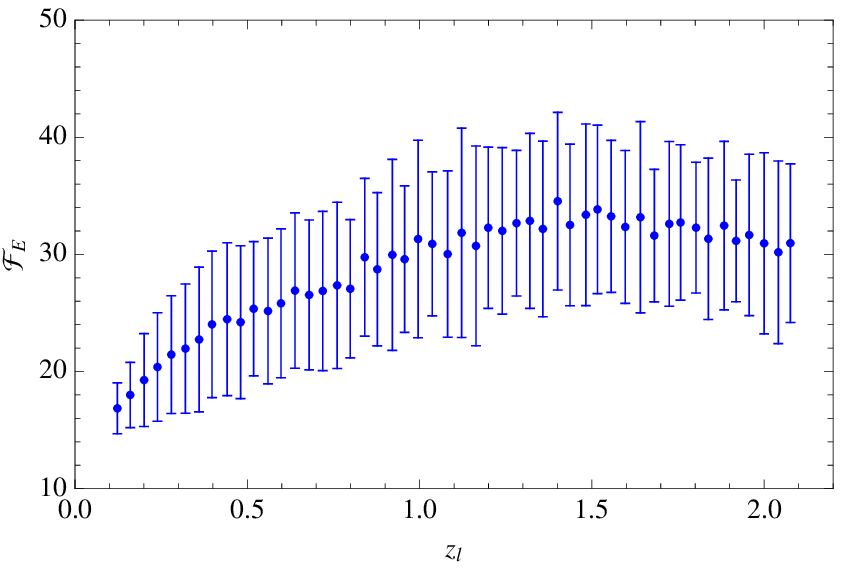}
\includegraphics[width=3.5cm]{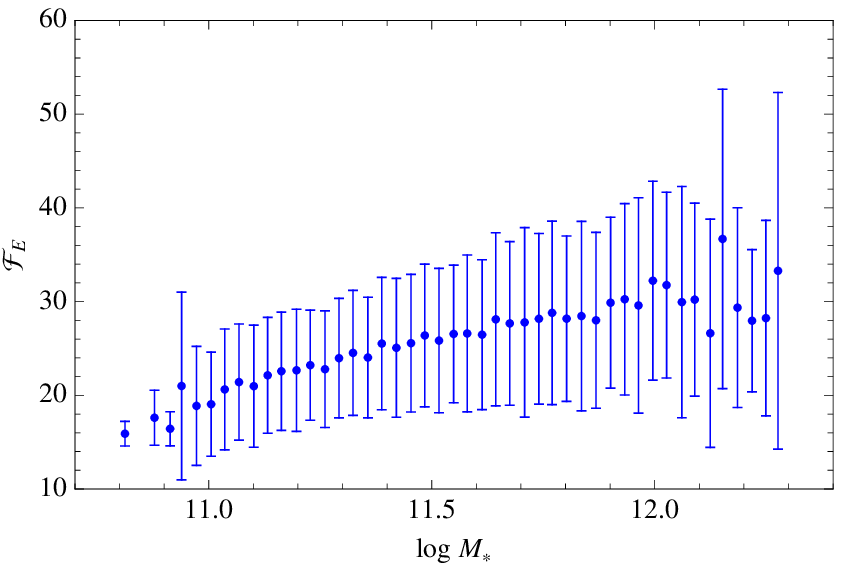}
\includegraphics[width=3.5cm]{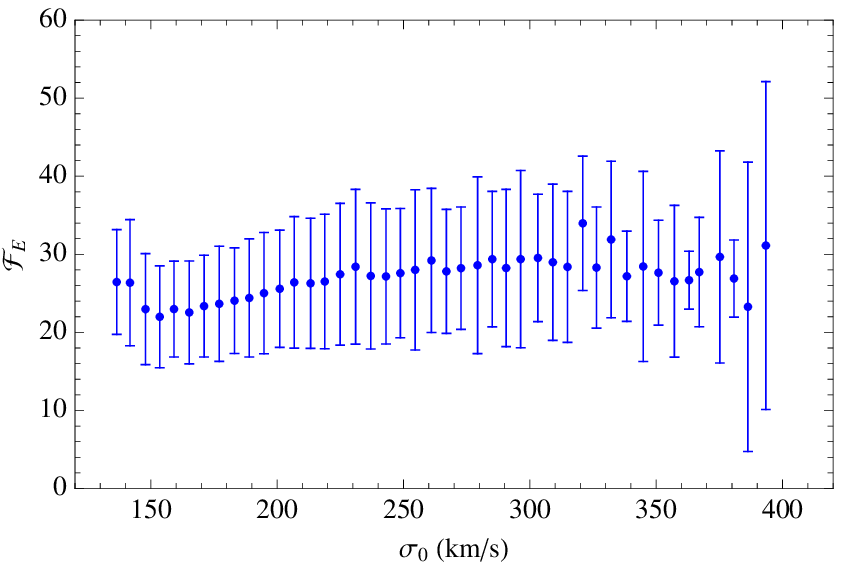}
\includegraphics[width=3.5cm]{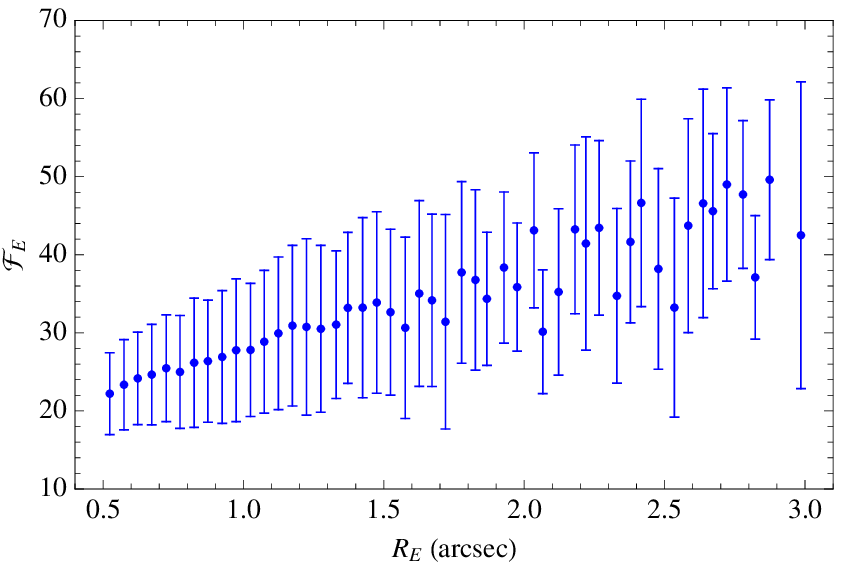}}
\resizebox{\hsize}{!}
{\includegraphics[width=3.5cm]{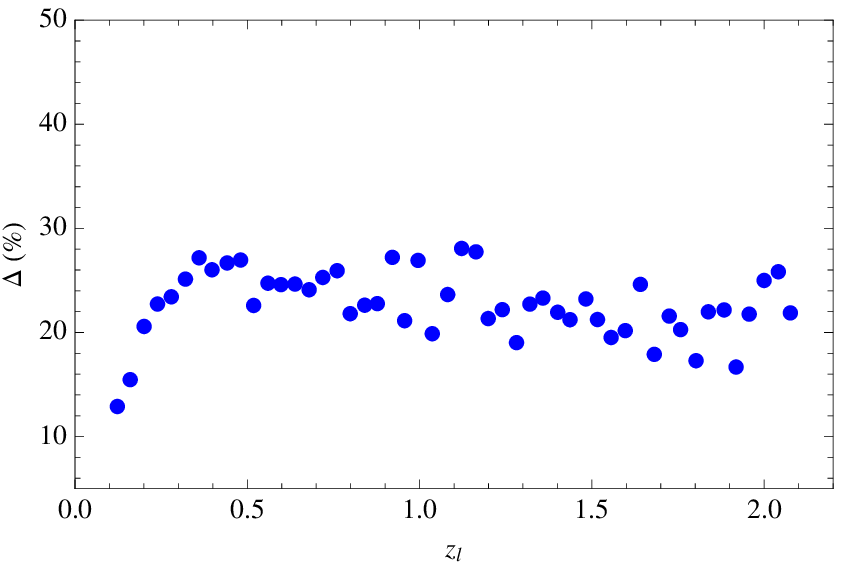}
\includegraphics[width=3.5cm]{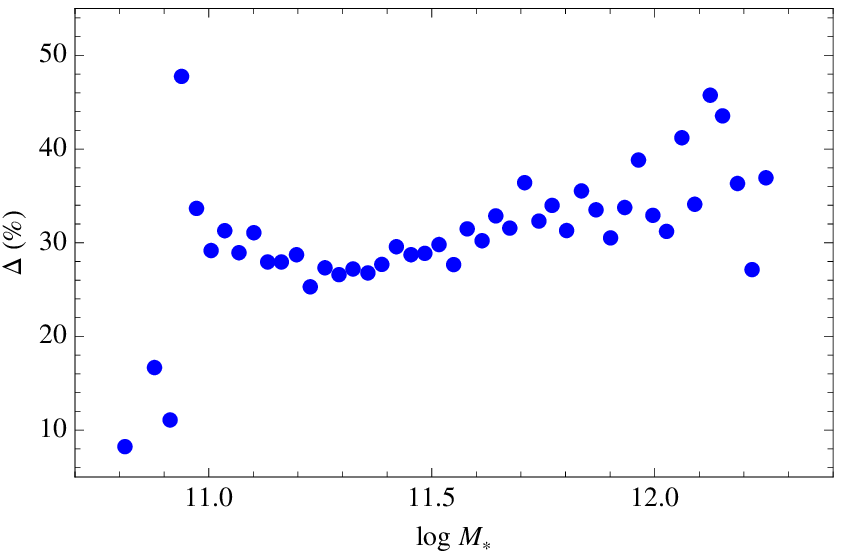}
\includegraphics[width=3.5cm]{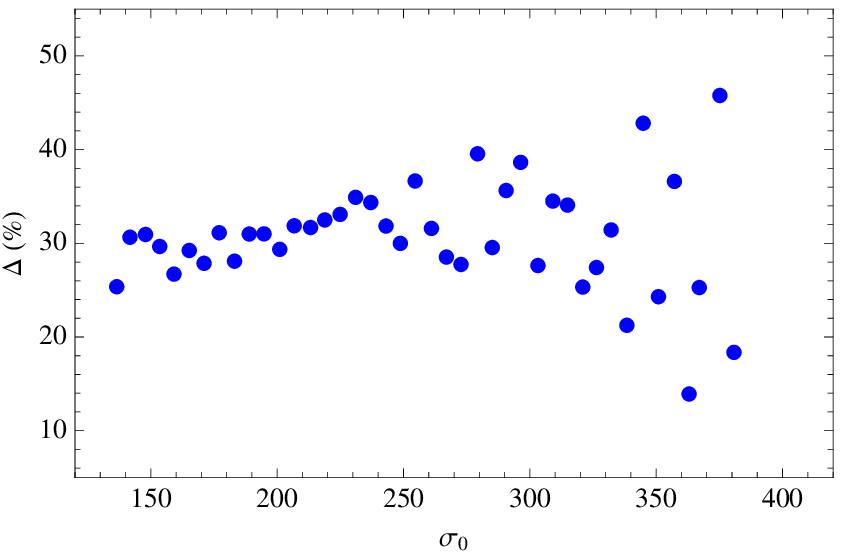}
\includegraphics[width=3.5cm]{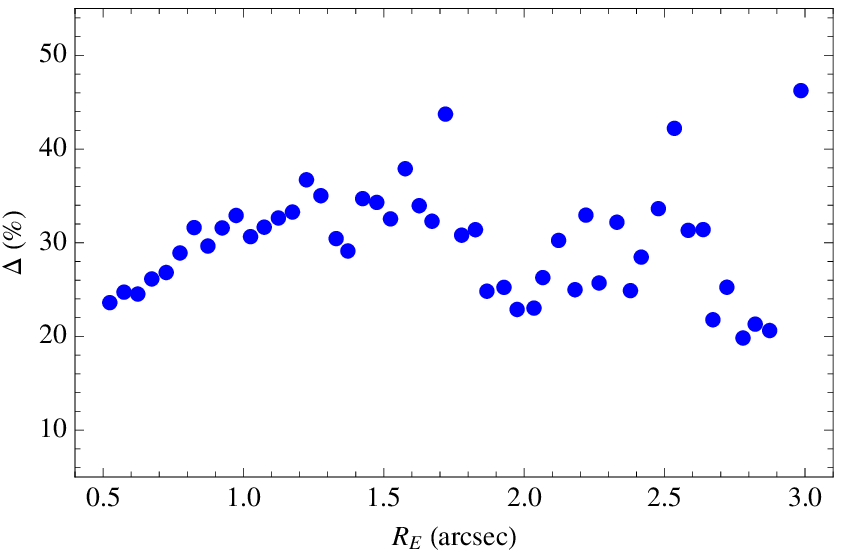}}
\caption{Nuisance parameter ${\cal{F}}_E$ and scatter $\Delta$ as functions of redshift $z_l$, stellar mass $\log{M_{\star}}$, central velocity dispersion $\sigma_0$, and Einstein radius $R_E$ for the 1000 lenses subsample. Results for the whole 10000 lenses sample are qualitatively the same.}
\label{fig: ffeplot}
\end{figure*}

\begin{equation}
\hat{{\cal{D}}} = D_{obs} \ \times \ \langle {\cal{F}}_E \rangle 
\label{eq: hatd}
\end{equation}
and consider the quantity 

\begin{displaymath}
\Delta = ({\cal{D}} - \hat{{\cal{D}}})/{\cal{D}} = ({\cal{F}}_E - \langle {\cal{F}}_E \rangle)/{\cal{F}}_E \ 
\end{displaymath}
as a measure of the relative error due to the distance ratio estimate. Since  $\langle {\cal{F}}_E \rangle$ is the average value over the lenses in a given bin (so that we can trace how $\Delta$ evolves with the bin centre), the question is now which parameter to use as binning quantity. 

As a first case, we consider the most obvious choice and divide the simulated sample in 50 redshift bins. It turns out that, while $\langle {\cal{F}}_E \rangle$ increases with $z_l$, its relative scatter only has a negligible variation, and that the root mean square value is $\Delta_{rms} \simeq 20\%$ independently of $z_l$. Moreover, we have verified that, for lenses in the same bin, $\Delta$ has a negligible correlation with the source redshift, the lens size and mass, and the halo model parameters.

We repeat this test, binning with respect to the stellar mass, central velocity dispersion, and Einstein radius. We find that we are unable to get satisfactory results. Indeed, binning with respect to $\log{M_{\star}}$ increases $\Delta_{rms}$ and introduces a significant increase with the stellar mass for $\log{M_{\star}} > 11.2$, with values going as large as $\sim 50\%$. Similarly large values are obtained when binning with respect to $\sigma_0$ although a definite trend is not found, while binning in $R_E$ introduces an increasing trend for $R_E < 1.5 \ {\rm arcsec}$ followed by a non monotonic variation leading to $\Delta_{rms} \simeq 30 - 40\%$ (but the low statistics at large $R_E$ values makes these numbers not fully reliable).

We therefore conclude that the estimator (\ref{eq: hatd}) is  reliable, once the lenses are divided in redshift bins. In this case, one can then rely on $\hat{{\cal{D}}}$ assuming that the nuisance parameter is the same for all the lenses in the same bin within a percentage scatter of $\sim 20\%$.

To further validate this conclusion, we have repeated the full analysis changing the assumptions in the simulation pipeline. We have indeeed considered a different $c_{vir}$\,-\,$M_{vir}$ relation either by adopting the one of \cite{MC11} or by changing the halo model from NFW to Einasto (1965, 1969) profile. Although the ${\cal{F}}_E$ values are different, the results on $\Delta_{rms}$ are qualitatively and quantitatively similar. In particular, we confirm the above conclusion that binning in lens redshift guarantees that $\Delta_{rms} \simeq 20\%$ independently of the bin centre and the  stellar and DM halo properties of the lenses.

\section{Fisher matrix forecast}

In order to investigate whether and under which conditions the method proposed works in constraining the cosmological parameters, we perform a Fisher matrix analysis. The elements of the Fisher matrix  are given by

\begin{displaymath}
F_{ij} = \left . \frac{\partial^2 {\cal{L}}}{\partial p_i \partial p_j} \right |_{{\bf p} = {\bf p}_{fid}}\,, 
\end{displaymath}
being  $p_i$ the i\,-\,th parameter and ${\bf p}_{fid}$ the fiducial values. To define the likelihood function ${\cal{L}}$ we split the sample in ${\cal{N}}_{b}$ bins and let ${\cal{N}}_k$ be the number of lenses in the $k$\,-\,th bin. We can compute the likelihood for lenses in this bin as

\begin{eqnarray}
{\cal{L}}_k & = & \frac{1}{2} \sum_{d = 1}^{{\cal{N}}_k}{\ln{(\sigma_d^2 + \sigma_E^2)}} \nonumber \\
 & + & \frac{1}{2} \sum_{d = 1}^{{\cal{N}}_k}{\left [ \frac{ {\cal{D}}_{obs,d} \langle {\cal{F}}_E \rangle_k - 
{\cal{D}}_{th}(z_{l,d}, z_{s,d}; {\bf p}_c)}{\sqrt{\sigma_d^2 + \sigma_E^2}} \right ]^2}
\label{eq: deflikek}
\end{eqnarray} 
where ${\bf p}_c$ is the set of cosmological parameters to be constrained, while $\langle {\cal{F}}_E \rangle_k$ is the average ${\cal{F}}_E$ value for the $k-th$ bin which we marginalize over. In order to compute the likelihood (and its derivatives) we need to assign the quantities entering Eq.(\ref{eq: deflikek}). The  ${\cal{D}}_{obs,d}$ may be easily computed for each lens using Eq.(\ref{eq: defdobs}); the errors $(\sigma_d, \sigma_E)$ deserve, on the other hand, some more words. First of all, $\sigma_d$ is the error on ${\cal{D}}_{obs,d} \langle {\cal{F}}_E \rangle_k$. Since we are assuming that $\langle {\cal{F}}_E \rangle_k$ is constant, we can then simply write $\sigma_d =  \langle {\cal{F}}_E \rangle_k \sigma_{obs}$, where a naive propagation of errors is sufficient to get

\begin{displaymath}
\sigma_{obs}/{\cal{D}}_{obs,d} = \sqrt{4 \varepsilon_E^2 + \varepsilon_{eff}^2 + 4 \varepsilon_0^2}\,.
\end{displaymath}
Here $(\varepsilon_E, \varepsilon_{eff}, \varepsilon_0)$ are the relative uncertainties on $(R_E, R_{eff}, \sigma_0)$, respectively. We will make the simplifying assumption that these quantities are the same for all the lenses in a given bin and explore different values for them (see later). The second term $\sigma_E$ is the systematic error induced by replacing the actual ${\cal{F}}_E$ with its average value and can be set as $\sigma_E = \Delta_{rms} \langle {\cal{F}}_E \rangle_k$. Based on the simulations described in the previous section, we will set $\Delta_{rms} = 0.2$, while $\langle {\cal{F}}_E \rangle_k$ is computed by binning the lenses according to the above chosen strategy.

Assuming that the lens redshift has been measured with a precision much smaller than the bin width, the total likelihood is simply\,:

\begin{displaymath}
{\cal{L}}({\bf p}_c, {\bf p_E}) = \prod_{k = 1}^{{\cal{N}}_b}{{\cal{L}}_k({\bf p}_c, \langle {\cal{F}}_E \rangle_k)}\,,
\end{displaymath}
where we collectively denote with ${\bf p}_E$ the set of $\langle {\cal{F}}_E \rangle_k$ values. As a consequence, the total Fisher matrix will be the sum of the Fisher matrices for each bin. It is convenient to first compute the Fisher matrix for the $k$\,-\,th bin, then marginalize over $\langle {\cal{F}}_E \rangle_k$, and finally sum the marginalized matrices to get ${\bf F}_c$, i.e.,  the total Fisher matrix for the cosmological parameters only.

\subsection{Input quantities}

In order to compute the Fisher matrix, some preliminary quantities must be set. First, we choose the fiducial cosmological model. We consider three different spatially flat models setting

\begin{displaymath}
(\Omega_M, w_0, w_a) = \left \{
\begin{array}{l}
(0.306, -1.0, 0.0) \\
(0.306, -0.95, 0.0) \\
(0.306, -0.90, -0.16) \\
\end{array}
\right .
\end{displaymath}
which we will refer to as $\Lambda$CDM, Quiessence, Thawing, respectively. Note that we have used the same matter density parameter for all three models, setting its value to the recent Planck estimate although one should change it according to the DE EoS parameters in order to get the best match to the data. However, we are here only interested in exploring realistic models so that we do not worry much about the exact value of the parameters. We nevertheless want to stress that these cases span the range of quintessence models still allowed by the data \citep{Lind15}. These parameters assign the dimensionless Hubble parameter $E(z) = H(z)/H_0$, reading

\begin{eqnarray}
E(z) & = & \Omega_M (1 + z)^3 \nonumber \\ 
 &  + & (1 - \Omega_M) (1 + z)^{3(1 + w_0 + w_a)} 
\exp{\left ( -\frac{w_a z}{1 + z} \right )} \ ,
\label{eq: hube}
\end{eqnarray}
which enters the distance ratio given by

\begin{equation}
{\cal{D}}(z_l, z_s) = \frac{D_{s}}{D_{ls}} = \frac{\chi(z_s)}{\chi(z_s) - \chi(z_l)}\,,
\label{eq: distratio}
\end{equation}
with 

\begin{equation}
\chi(z) = \int_{0}^{z}{E^{-1}(\zeta) d\zeta}
\label{eq: comdist}
\end{equation}
the comoving distance. Note that Eq.(\ref{eq: distratio}) only holds for spatially flat models. We also stress that the Hubble constant $H_0$ drops out from the ratio of distances thus reducing the number of cosmological parameters.

A key role in determining the accuracy of the constraints is played by the errors on the observable quantities. These enters Eq.(\ref{eq: deflikek}) through the statistical term $\sigma_{obs}$ and the systematic one $\sigma_E$. As already said above, we set $\sigma_E = \Delta_{rms} \langle {\cal{F}}_E \rangle_k$, since this is the scatter in the distance ratio introduced by the approximation done when using the estimator $\hat{{\cal{D}}}$ instead of the correct one in Eq.(\ref{eq: testeq}). On the other hand, the statistical uncertainty $\sigma_{obs}$ depends on the relative errors on $(R_E, R_{eff}, \sigma_0)$. We will set $\varepsilon_{eff} = 0.015$ as typically found fitting HST\,-\,like data on lens surface brightness profile, while we try three different configurations for the uncertainties on the Einstein radius and central velocity dispersion, namely

\begin{displaymath}
(\varepsilon_E, \varepsilon_0) = 
(0.05, 0.10) \ , \ (0.01, 0.10) \ , \ (0.01, 0.05) \ .
\end{displaymath}
As a final ingredient, we need to set the total number of lenses and the binning. We consider two cases. First, we take 1000 lenses splitted in 20 equally spaced bins over the redshift range $(0.1, 2.1)$, while secondly we take 10000 lenses splitted in 50 bins over the same redshift range as improved sample.

\begin{table}
\caption{Marginalized constraints on $(\Omega_M, w_0)$ for $\Lambda$CDM, Quiessence and Tahwing models (top, centre and bottom part of the table, respectively) for different sample and error configurations. The labels $1k$ and $10k$ refer to the sample containing 1000 and 10000 lenses, respectively. We remind the reader that we set $\varepsilon_{eff} = 0.015$ for all the cases.}
\centering
\begin{tabular}{cccccc}
\hline
$\varepsilon_E$ & $\varepsilon_0$ & $\sigma_{1k}(\Omega_M)$ & $\sigma_{1k}(w_0)$  & $\sigma_{10k}(\Omega_M)$ & $\sigma_{10k}(w_0)$ \\
\hline
0.05 & 0.10 & 0.026 & 0.195 & 0.008 & 0.060 \\
0.01 & 0.10 & 0.024 & 0.184 & 0.008 & 0.056 \\
0.01 & 0.05 & 0.019 & 0.140 & 0.006 & 0.042 \\
\hline
0.05 & 0.10 & 0.027 & 0.191 & 0.008 & 0.058 \\
0.01 & 0.10 & 0.026 & 0.180 & 0.008 & 0.055 \\
0.01 & 0.05 & 0.020 & 0.137 & 0.006 & 0.042 \\
\hline
0.05 & 0.10 & 0.026 & 0.183 & 0.008 & 0.056 \\
0.01 & 0.10 & 0.025 & 0.172 & 0.008 & 0.052 \\
0.01 & 0.05 & 0.019 & 0.131 & 0.006 & 0.040 \\
\hline
\end{tabular}
\label{tab: fishmatrescst}
\end{table}

\subsection{Results}

Table\,\ref{tab: fishmatrescst} and \ref{tab: fishmatresvar} summarizes the results of the Fisher matrix forecast for different models, sample and error configurations. First, we consider the simplifying assumption that $w_a$ is known\footnote{From the point of view of the Fisher matrix, this means we delete the row and columns corresponding to $w_a$ \citep{Coe}.}. While for $\Lambda$CDM and Quiessence models this means that, in a hypothetical fit, we force $w_a = 0$, in the case of Thawing model, one has to set $w_a = -1.58 (1 + w_0)$ in order to get the correct behaviour \citep{Lind15}.

\begin{figure*}
\resizebox{\hsize}{!}
{\includegraphics[width=4.5cm]{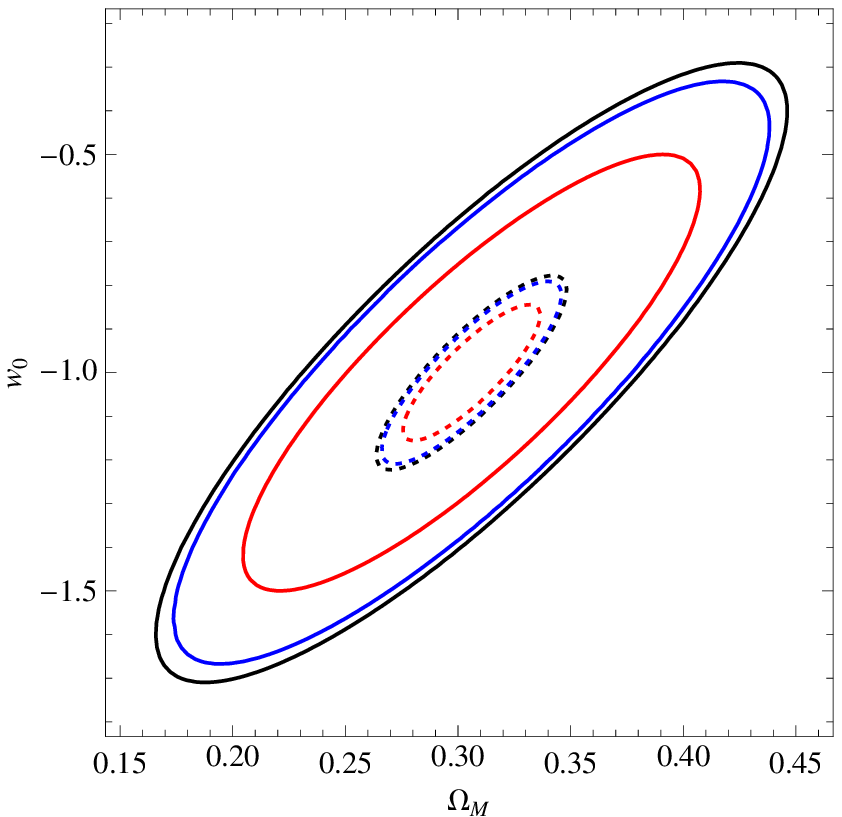}
\includegraphics[width=4.5cm]{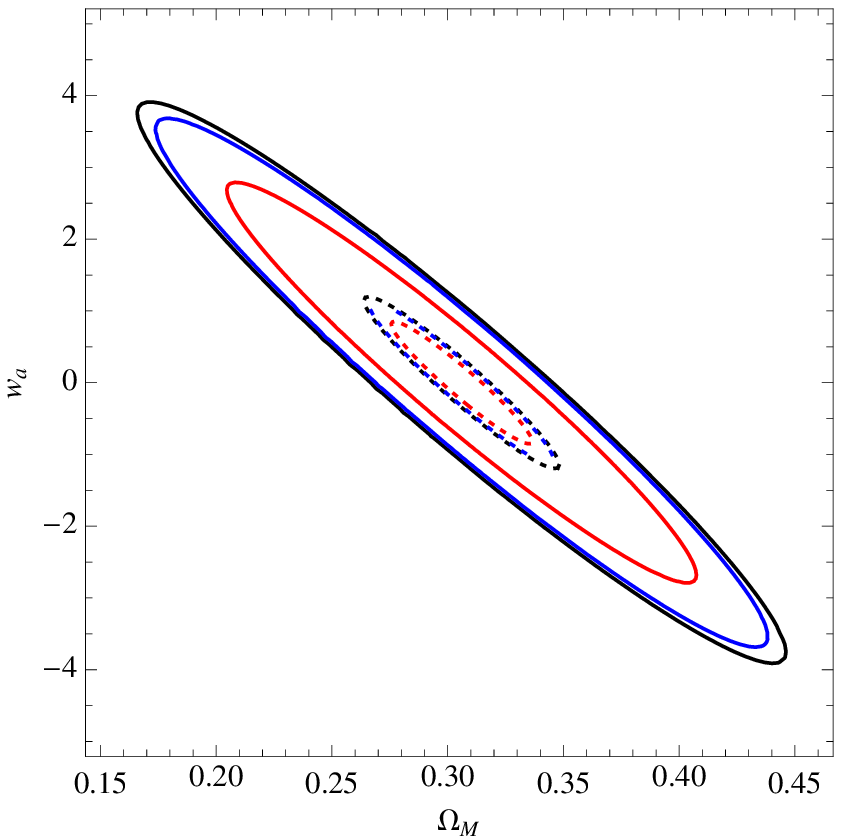}
\includegraphics[width=4.5cm]{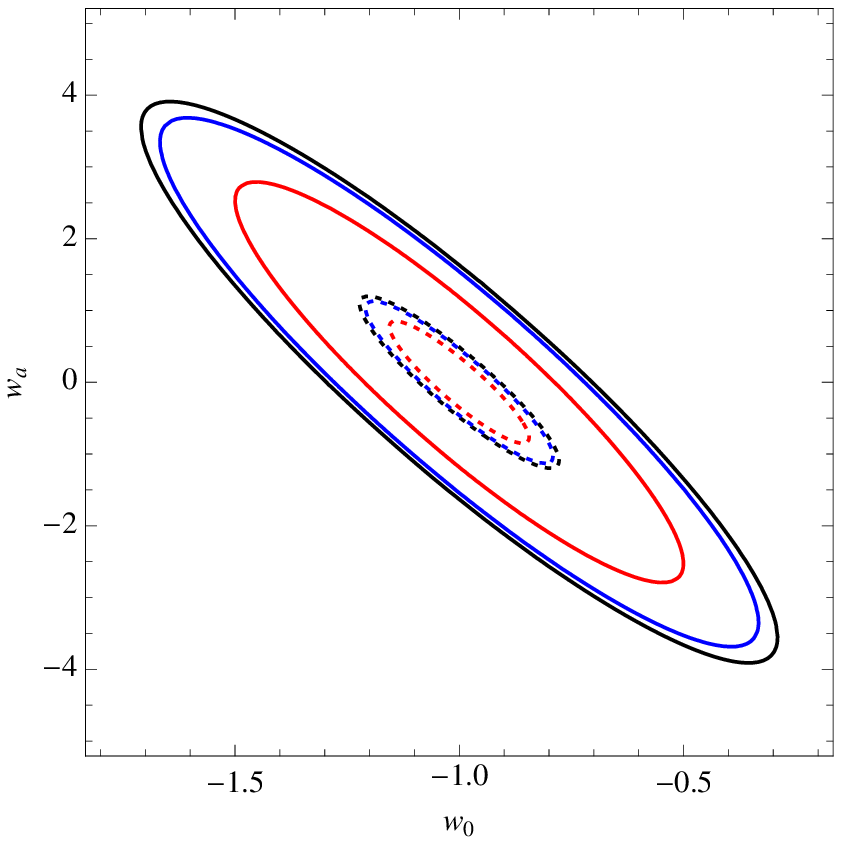}}
\caption{$68\%$ contour plots for the fiducial $\Lambda$CDM model as inferred from 1000 (solid) and 10000 (dotted) lenses samples under the three different assumptions on the errors (black, blue and red) in the same order as in the text. }
\label{fig: contplot}
\end{figure*}

The numbers in Table\,\ref{tab: fishmatrescst} convincingly show that the distance ratio method is quite effective at constraining the cosmological parameters $(\Omega_M, w_0)$ if a prior on $w_a$ is set. No matter which model is taken as fiducial, we find that 1000 lenses are sufficinet to get $\sigma(\Omega_M)/\Omega_M \sim 10\%$ and $\sigma(w_0)/w_0 \sim 13 - 20\%$. Increasing the sample by an order of magnitude roughly halves these numbers pointing at a possible saturation of the method accuracy. For fixed model and lens sample, it turns out that reducing the error on $\sigma_0$ is a better strategy to improve the constraints since it is more effective at narrow down the error on $w_0$. The constraints are basically the same for the three models, although those for the Thawing case are consistently smaller. This can be qualitatively explained noting that, in this case, $w_0$ has a larger impact on the distance ratio thanks to the presence of the exponential term in the dimensionless Hubble parameter (which drops out when $w_a = 0$ is used).

Although encouraging, these results are based on the strong prior that we know how to set $w_a$. In a realistic application, however, one only assumes that the DE EoS is given by the CPL approximation and fits for the three parameters $(\Omega_M, w_0, w_a)$. The constraints we get are summarized in Table\,\ref{tab: fishmatresvar} and show a significant degradation with respect to the case with $w_a$ fixed. While the trends with $(\varepsilon_E, \varepsilon_0)$ are qualitatively the same, the number of lenses becomes now of primary importance. Indeed, 10000 rather than 1000 lenses are now needed to get $\sigma(\Omega_M)/\Omega_M \sim 10\%$ and $\sigma(\Omega_M)/\Omega_M \sim 15\%$, while only weak constraints can be put on $w_a$. This result can be easily understood looking at the pivot redshift $z_p$ defined as the value of $z$ where the errors are uncorrelated. Not surprisingly, we find that $z_p \simeq 0.20$ no matter which error configuration, lens sample or fiducial model is used. This is an obvious consequence of the lens redshift distribution which indeed has a median value close to the pivot redshift. As a result, the DE EoS is best constrained at $z_p$ with the error on $w_p = w(z_p)$ turning out to be\footnote{Hereafter, we only give the results for the fiducial $\Lambda$CDM, but they are quite similar for other models.}

\begin{displaymath}
\sigma(w_p) = (0.081, 0.076, 0.059)
\end{displaymath}
from 1000 lenses in the three error configurations and 

\begin{displaymath}
\sigma(w_p) = (0.024, 0.023, 0.018)
\end{displaymath}
from 10000 lenses. These results clearly show that the distance ratio method is better suited at probing the low redshift behaviour of the DE EoS so that its redshift evolution is hardly probed thus explaining the weak constraints on $w_a$. 

It is interesting to look at the FoM of the method. We get ${\rm FoM} = (2.0, 2.2, 3.9)$ when using 1000 lenses, while it is ${\rm FoM} = (21.2, 23.9, 41.9)$ if 10000 lenses are used. It is evident that interesting values can only be obtained reducing as much as possible the error on $\sigma_0$ and increasing the lens sample. We, however, note that the orientation of the $(w_0, w_a)$ ellipses (see Fig,\,\ref{fig: contplot}), is quite different from other classical methods such as SNeIa and CMBR thanks to the different pivot redshift. It is therefore likely that a combination of the distance ratio method with these other probes helps breaking degeneracies thus boosting the total FoM.

\begin{table}
\caption{Same as Table\,\ref{tab: fishmatrescst} but for $(\Omega_M, w_0, w_a)$.}
\resizebox{8.5cm}{!}{
\begin{tabular}{cccccccc}
\hline
$\varepsilon_E$ & $\varepsilon_0$ & $\sigma_{1k}(\Omega_M)$ & $\sigma_{1k}(w_0)$  & $\sigma_{1k}(w_a)$ & $\sigma_{10k}(\Omega_M)$ & $\sigma_{10k}(w_0)$ & $\sigma_{10k}(w_a)$ \\
\hline
0.05 & 0.10 & 0.092 & 0.47 & 2.58 & 0.028 & 0.15 & 0.79 \\
0.01 & 0.10 & 0.087 & 0.44 & 2.43 & 0.026 & 0.14 & 0.74 \\
0.01 & 0.05 & 0.067 & 0.33 & 1.84 & 0.020 & 0.10 & 0.56 \\
\hline
0.05 & 0.10 & 0.100 & 0.44 & 2.48 & 0.030 & 0.14 & 0.76 \\
0.01 & 0.10 & 0.095 & 0.42 & 2.34 & 0.029 & 0.13 & 0.72 \\
0.01 & 0.05 & 0.073 & 0.31 & 1.77 & 0.022 & 0.10 & 0.54 \\
\hline
0.05 & 0.10 & 0.093 & 0.44 & 2.38 & 0.028 & 0.14 & 0.73 \\
0.01 & 0.10 & 0.088 & 0.42 & 2.24 & 0.026 & 0.13 & 0.69 \\
0.01 & 0.05 & 0.067 & 0.31 & 1.69 & 0.020 & 0.10 & 0.52 \\
\hline
\end{tabular}}
\label{tab: fishmatresvar}
\end{table}

\section{Conclusions}

Einstein rings have always attracted a lot of attention not only for their spectacular beauty, but also as a probe of the dark matter content of lens galaxies. The measurement of both the Einstein radius and central velocity dispersion allows to strengthen the constraints on the halo density profile, but can also be used as a way to estimate the ratio between the distance to the source and that between lens and source and hence probe the cosmic expansion. Motivated by the previous literature results, we have proposed a novel approach which, although based on the same idea, ameliorates the estimate of the distance ratio. On one hand, we have adopted a realistic two components model for the lens galaxy taking care of both the stellar and DM contributions. This led us to a general formula which permits to estimate the distance ratio ${\cal{D}} = D_{s}/D_{ls}$ as a function of observable quantities and a nuisance parameter ${\cal{F}}_E$ depending on the details of the halo model and the lens parameters. The analytical formula we derive for ${\cal{F}}_E$ has allowed us to show that its value is approximately constant for lenses belonging to the same redshift bin, so that we finally get an approximate estimator which helps us to define a likelihood function for the estimate of cosmological parameters. It is worth stressing that the method proposed does not depend on the particular halo model adopted or on the details of the $M_{\star}/M_{vir}$\,-\,$M_{\star}$ and $c_{vir}$\,-\,$M_{vir}$ relations adopted. This is a further improvement with respect to the standard approach which, on the constrary, postulates that all lenses may be described by the singular isothermal sphere.

A well founded and movitated method to constrain cosmological parameters should be useless if the numbers of tracers and/or the requirements on the uncertainties on the observable quantities are too demanding. To investigate this, we have carried on a Fisher matrix analysis changing both the number of lenses and the relative errors on $(R_E, \varepsilon_0)$. It turns out that a survey measuring $(R_E, \sigma_0)$ with $(1, 5)\%$ accuracy for 10000 lenses can get ${\rm FoM} \sim 40$, while $\sim 10\%$ constraints on $(\Omega_M, w_0)$ can be obtained with 1000 lenses only if $w_a$ is fixed. It is therefore worth wondering whether these requirements are realistic. To this end, we note that present day lenses already achieved $\varepsilon_E \sim 5\%$ and $\varepsilon_0 \sim 10\%$, so that it is not unrealistic to guess that higher quality images (as can be obtained from future large telescopes and satellite missions) and improved spectrograph can easily achieve our requirements. More demanding is the constraint on the total number of lenses. As an example, however, one can note that the Euclid mission \citep{RB} is expected to observe up to $170000$ strong lenses \citep{Col15}, so that asking that both $R_E$ and $\sigma_0$ are measured for less than $10\%$ of them is not a too demanding requirement. 

Although the present day lens samples are far from the numbers explored in our Fisher matrix analysis, it could be nevertheless interesting to apply the proposed method to real strong lensing data. \cite{Cao15} have recently assembled a catalog of 118 strong lensing systems and used it to constrain the dark energy equation of state under the usual assumption of isothermal sphere model. We plan to repeat their analysis using our improved approach in order to investigate to which extent the constraints on the cosmological parameters depend on the assumed lens model. Actually, given the small statistics, it is expected that the large errors could prevent a conclusive answer, but such a study can provide a comparison benchmark for an analysis based on simulated samples.

It is worth noting that the results are always somewhat dependent on the lens redshift distribution. Here, we have simulated lens samples in such a way that they grossly reproduce the main features of the SLACS dataset since this is the most used catalog at the moment. However, future datasets can have a different redshift distribution, possibly changing the method FoM. Although a more detailed analysis will be carried on in a future work, we can qualitatively anticipate that the pivot redshift will likely be close to the median survey redshift. It is therefore possible that combining different samples with different median redshifts could help better tracing the DE EoS over a larger redshift range thus increasing the FoM. Note that we are here implicitly assuming that the median redshift of the full lens sample is the same as the median redshift of the subsample with measured values of $(z_l, z_s, R_E, \sigma_0)$, which can also not be the case because of selection effects. On the contrary, since not all the lenses contribute to the FoM in  the same way, it is possible that an optimal strategy can be worked out maximizing the scientific return of the companion spectroscopic survey needed to measure the lenses velocity dispersions. To this end, one can follow the same methodology presented, for time delay distances,  in \cite{L15}.

As a final remark, we stress that, even if suboptimal conditions in terms of number of lenses and/or accuracy in the measurement of $(R_E, \sigma_0)$ are achieved, the distance ratio method we have proposed can be efficiently combined with other probes helping to break degeneracies and hence boosting the total FoM. Should the present results be confirmed by a more detailed analysis of the lens samples properties, new light on dark energy will be shed by the strong regime of gravitational lensing.

\section*{Acknowledgments}

VFC is funded by Italian Space Agency (ASI) through contract Euclid\,-\,IC (I/031/10/0) and acknowledges financial contribution from the agreement ASI/INAF/I/023/12/0.

\end{document}